\documentclass{ceurart}

\usepackage{amsmath}
\usepackage{amssymb}
\usepackage{booktabs}
\usepackage{graphicx}
\usepackage{algorithm}
\usepackage{algorithmic}
\usepackage{multirow}

\graphicspath{{figures/}}

\begin{document}

\copyrightyear{2026}
\copyrightclause{Copyright for this paper by its authors.
  Use permitted under Creative Commons License Attribution 4.0
  International (CC BY 4.0).}
\conference{ECOM'26: SIGIR Workshop on eCommerce, Jul 24, 2026, Melbourne, Australia}

\title{AutoRelAnnotator: Calibrated Model Cascades for Cost-Efficient Relevance Evaluation in Sponsored Search}

\author[1]{Md Omar Faruk Rokon}[%
  email=mdomarfaruk.rokon@walmart.com,
]
\cormark[1]
\author[1]{Shasvat Desai}[%
  email=shasvat.desai@walmart.com,
]
\cormark[1]
\author[1]{Hong Yao}[%
  email=hong.yao0@walmart.com,
]
\author[1]{Kuang-chih Lee}[%
  email=kuang-chih.lee@walmart.com,
]

\address[1]{Walmart Global Tech, Sunnyvale, CA, USA}

\cortext[1]{Corresponding author.}

\begin{abstract}
How can we generate high-quality relevance annotations at scale without the cost and delays of human labeling?
Relevance annotations are the backbone of search ranking systems which is needed for training data preparation, NDCG evaluation, and root cause analysis. However, human annotation is slow and off-the-shelf LLMs suffer from accuracy on domain-specific tasks.
We propose a \textbf{calibrated model cascade}, a systematic approach for cost-efficient offline relevance annotation by routing queries through progressively larger fine-tuned classifiers.
Our central insight is that \textbf{accuracy and cost are orthogonal optimizations}: domain-specific fine-tuning drives accuracy, cascading drives cost, and per-class isotonic calibration adds a small but reliable gain on top.
Our contribution is threefold:
(a) we decompose the gains and show that fine-tuning contributes ${\sim}20$ accuracy points while cascading is approximately accuracy-neutral but halves compute cost,
(b) we introduce per-class isotonic calibration as one component of the cascade, contributing a small but statistically significant gain (+0.6 points over the strongest calibration baseline), and
(c) we validate the system in production across six offline use cases, processing 150M+ annotations and enabling faster experimentation cycles.
Our work is a building block for scalable, high-quality offline annotation pipelines in search and advertising systems.
\end{abstract}

\begin{keywords}
Relevance Annotation \sep
Model Cascades \sep
Confidence Calibration \sep
Sponsored Search \sep
LLM Classification \sep
E-Commerce Search
\end{keywords}

\maketitle

\section{Introduction}

How can a search team evaluate ranking quality, diagnose relevance failures, or prepare training data---when human annotation takes days and costs hundreds of thousands of dollars?
This is not a hypothetical problem.
In our production environment, a team of 3 trained annotators produces approximately 3,000 labels per day at \$0.50 per annotation, with 3--5 business day turnaround for batches exceeding 10,000 labels.
These constraints bottleneck virtually every stage of the search ML lifecycle:
(a) \textbf{training data preparation}: annotating query-product pairs for ranking model training,
(b) \textbf{NDCG evaluation}: assessing search quality using graded relevance judgments,
(c) \textbf{root cause analysis}: rapidly annotating suspect queries when production metrics degrade,
(d) \textbf{guardrail calibration}: determining thresholds for production safety systems, and
(e) \textbf{feature evaluation}: measuring the impact of A/B tests on relevance.
None of these use cases require real-time inference. However, these require high accuracy at manageable cost.
This offline setting is crucial. Unlike online serving where latency is paramount, batch annotation allows us to run multiple models, use ensemble fallbacks, and invest in calibration. These are impractical at serving time but highly effective for annotation quality.

\begin{figure}[t]
\centering
\includegraphics[width=0.6\columnwidth]{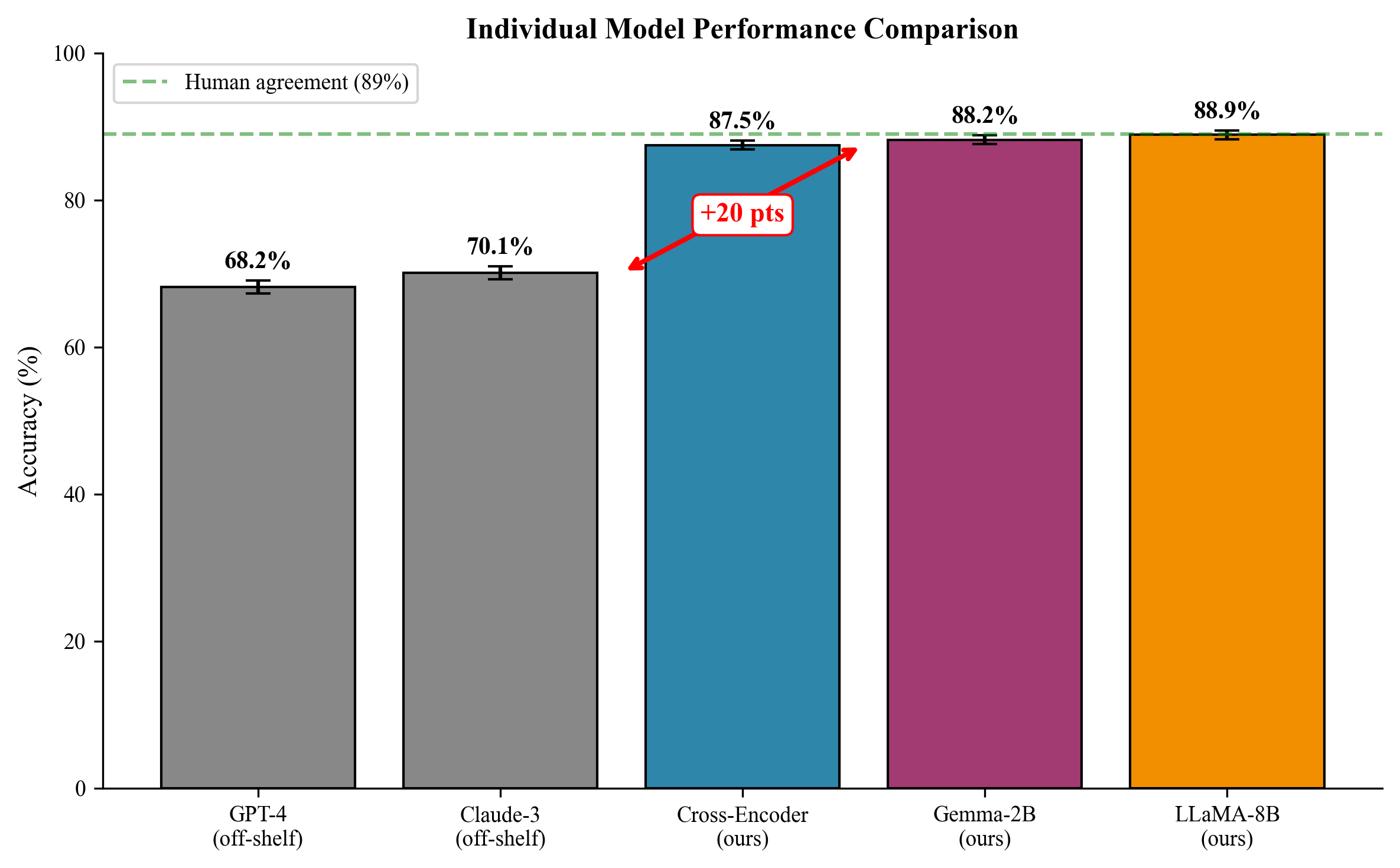}
\caption{Accuracy-cost tradeoff: fine-tuning improves accuracy, while cascading reduces compute cost.}
\label{fig:pareto}
\end{figure}

Can off-the-shelf LLMs fill this gap?
Unfortunately, no.
GPT-4 achieves only 68.2\% and Claude-3 only 70.1\% accuracy on our 5-class relevance task which are sub par compared to the 89\% inter-annotator agreement we observe.
LLM cascade methods \cite{chen2023frugalgpt,kolawole2024agreement,ong2024routellm} reduce costs by routing among such models, but inherit their accuracy ceiling which is a fundamental limitation, since no routing policy can substantially exceed the accuracy of its constituent models \cite{jitkrittum2023when}.

We take a fundamentally different approach.
We propose a \textbf{calibrated model cascade}, a systematic approach for offline relevance annotation that decouples two orthogonal problems:
(a) \textit{accuracy improvement} through domain-specific fine-tuning, and
(b) \textit{cost reduction} through calibrated cascading.
Rather than routing among low performing models, we first fine-tune classifiers achieving higher accuracy individually, then apply per-class isotonic calibration to route queries through a three-model cascade (Cross-Encoder $\rightarrow$ Gemma-2B $\rightarrow$ LLaMA-8B).
Figure~\ref{fig:pareto} illustrates these orthogonal gains.

\textbf{Contribution.}
The contribution of this work is threefold:

\noindent\textbf{a.} We show that \textit{accuracy and cost are orthogonal optimizations} for offline relevance annotation: fine-tuning contributes a 20+ point accuracy gain while cascading is approximately accuracy-neutral but halves compute cost.

\noindent\textbf{b.} We introduce per-class isotonic calibration as one component of the cascade, contributing a small but statistically significant gain (+0.6 points over the strongest calibration baseline, $p < 0.05$).

\noindent\textbf{c.} We validate the system across six offline use cases, processing over 150 million annotations and reducing turnaround from 5 days to 1--3 hours.

\noindent\textbf{Our work in perspective.}
Our approach can be seen as a first step towards fully automated offline annotation pipelines.
The cascade is modular where each model can be independently updated without affecting routing. It can be a practical building block for any organization needing large-scale relevance annotations.

\section{Related Work}

\paragraph{LLM Cascades and Routing.}
Cascade architectures route inputs through progressively more expensive models.
FrugalGPT \cite{chen2023frugalgpt} learns a DistilBERT scoring function to select among LLM APIs, achieving up to 98\% cost reduction.
ABC \cite{kolawole2024agreement} uses ensemble agreement as a training-free deferral signal for 2--25$\times$ cost savings; its extension to open-ended generation via semantic agreement appeared at EMNLP 2025 \cite{soiffer2025semantic}.
RouteLLM \cite{ong2024routellm} trains routers on preference data for binary model selection.
Hybrid LLM \cite{ding2024hybrid} learns a quality predictor for routing between two models.
Zellinger and Thomson \cite{zellinger2025rational} provide a probabilistic framework for threshold optimization using Markov-copula models.
Dekoninck et al.\ \cite{dekoninck2024unified} derive optimal strategies for unified routing and cascading.
C3PO \cite{valkanas2025c3po} applies conformal prediction for probabilistic cost bounds.

While these cascade methods achieve impressive cost reductions, our work differs in two ways:
(a) we fine-tune classifiers before cascading, decoupling accuracy gains (from fine-tuning) from cost gains (from cascading), rather than routing among off-the-shelf models that inherit their accuracy limitations on domain-specific tasks, and
(b) we apply per-class calibration for routing decisions.
Three recent papers move closer to our approach.
CascadeBERT \cite{li2021cascadebert} cascades multiple BERT variants with calibrated confidence which is our closest pre-LLM precursor, which we extend to heterogeneous LLM architectures.
Cascade-Aware Training \cite{wang2024cascade} fine-tunes the smallest model with downstream awareness, but targets generative tasks only.
GATEKEEPER \cite{rabanser2025gatekeeper} modifies the training loss for calibration-aware deferral. Our approach differs in using standard fine-tuning plus \textit{post-hoc} per-class isotonic calibration, maintaining modularity.

\paragraph{Confidence Calibration.}
Neural networks produce miscalibrated confidence scores \cite{guo2017calibration}.
Post-hoc methods include temperature scaling, Platt scaling \cite{platt1999probabilistic}, histogram binning \cite{zadrozny2001obtaining}, and isotonic regression \cite{zadrozny2002transforming}.
Nixon et al.\ \cite{nixon2019measuring} demonstrate that class-conditional miscalibration is far worse than global ECE suggests.
Kull et al.\ \cite{kull2019beyond} formalize classwise calibration and propose Dirichlet calibration as an alternative. We choose per-class isotonic regression for its nonparametric flexibility.
Jitkrittum et al.\ \cite{jitkrittum2023when} provide theoretical conditions under which confidence-based cascade deferral succeeds, noting that well-calibrated scores are essential.

\paragraph{Relevance Classification.}
Cross-encoders \cite{nogueira2019passage,reimers2019sentence} jointly encode query-document pairs for relevance scoring. DeBERTa-v3 cross-encoders arguably match models 11$\times$ larger on reranking tasks \cite{dejean2024cross}.
LoRA \cite{hu2022lora} enables efficient fine-tuning of LLaMA \cite{touvron2023llama} and Gemma \cite{gemma2024}.
LLMs as annotators have gained attention \cite{gilardi2023chatgpt,he2024annollm}, but studies consistently find that generic LLMs underperform on domain-specific tasks \cite{wang2024pandalm}.

\section{Our Approach}

Why do existing cascades plateau at 68--72\% accuracy on domain-specific tasks?
The answer is straightforward: they route among off-the-shelf models whose individual accuracy is limited \cite{chen2023frugalgpt,kolawole2024agreement}.
No matter how sophisticated the routing, the cascade cannot substantially exceed the accuracy of its constituent models.

\begin{figure*}[h]
\centering
\includegraphics[width=0.9\textwidth]{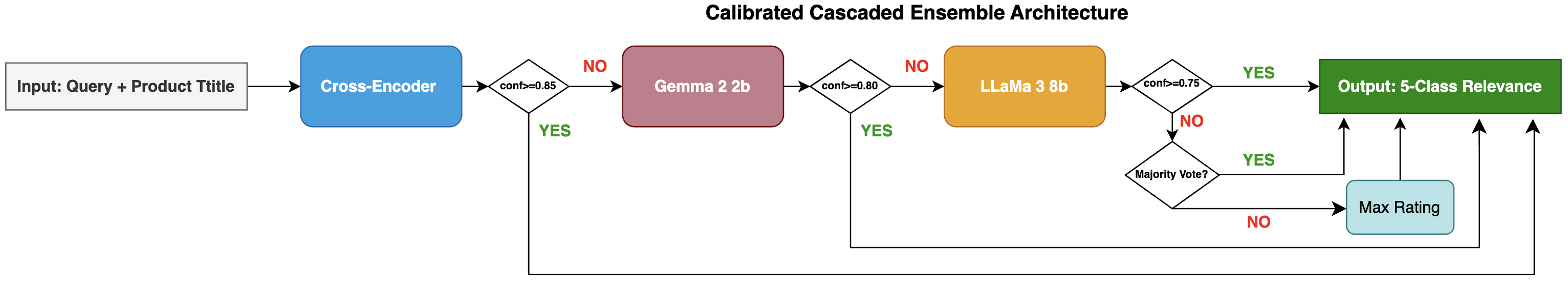}
\caption{Offline cascade with calibrated deferral at each stage and an ensemble fallback for ambiguous cases.}
\label{fig:architecture}
\end{figure*}

Our key insight is to \textbf{decouple accuracy improvement from cost reduction}:
(a) first, we invest in domain-specific fine-tuning to create high-accuracy classifiers (87--89\%), and
(b) then, we apply calibrated cascading to reduce computational cost while preserving this accuracy.

\subsection{System Architecture}

Our cascade consists of three domain-specific fine-tuned classifiers ordered by cost.
Figure~\ref{fig:architecture} illustrates the full pipeline.

\noindent\textbf{Cross-Encoder} (DeBERTa-v3-base, 184M params): Processes query-title pairs in ${\sim}$5ms. We fine-tuned with a 5-class classification head.
\textbf{Gemma-2B}: We fine-tuned with LoRA ($r{=}16, \alpha{=}16$). Inference ${\sim}$50ms.
\textbf{LLaMA-8B} (Meta LLaMA-3-8B): Our largest cascade model with similar LoRA fine-tuning. Inference ${\sim}$200ms.
All latency measured on NVIDIA A100 (40GB), batch size 64, FP16.

\subsection{Domain-Specific Fine-Tuning}

How do we obtain reliable confidence scores?
Unlike generative approaches requiring regex parsing of free-form text, our models use classification heads producing well-defined confidence:
\begin{equation}
p(y|q,t) = \text{softmax}(\mathbf{W} \cdot \mathbf{h}_{\text{pool}} + \mathbf{b})
\end{equation}
where $\mathbf{h}_{\text{pool}}$ is the pooled sequence representation, $\mathbf{W} \in \mathbb{R}^{5 \times d}$, and $y \in \{0,1,2,3,4\}$.
For the Cross-Encoder, $\mathbf{h}_{\text{pool}}$ is the final [CLS] hidden state.
For Gemma and LLaMA, $\mathbf{h}_{\text{pool}}$ is the mean-pooled final hidden state across tokens; LoRA targets attention projections and the classifier head trains end-to-end.
Training uses cross-entropy loss with early stopping on validation loss.

\subsection{Per-Class Isotonic Calibration}

Why not use raw confidence for routing?
Raw confidences are miscalibrated \cite{guo2017calibration}, and prior cascades either ignore calibration \cite{kolawole2024agreement} or apply global calibration that treats all classes identically \cite{chen2023frugalgpt}.
We observe that confidence distributions vary across classes: models are overconfident on extreme classes (0, 4) and underconfident on middle classes (1, 2, 3) which is consistent with Nixon et al.\ \cite{nixon2019measuring}, who show standard ECE underestimates class-conditional miscalibration by 3--5$\times$.

\begin{figure}[t]
\centering
\includegraphics[width=0.8\columnwidth]{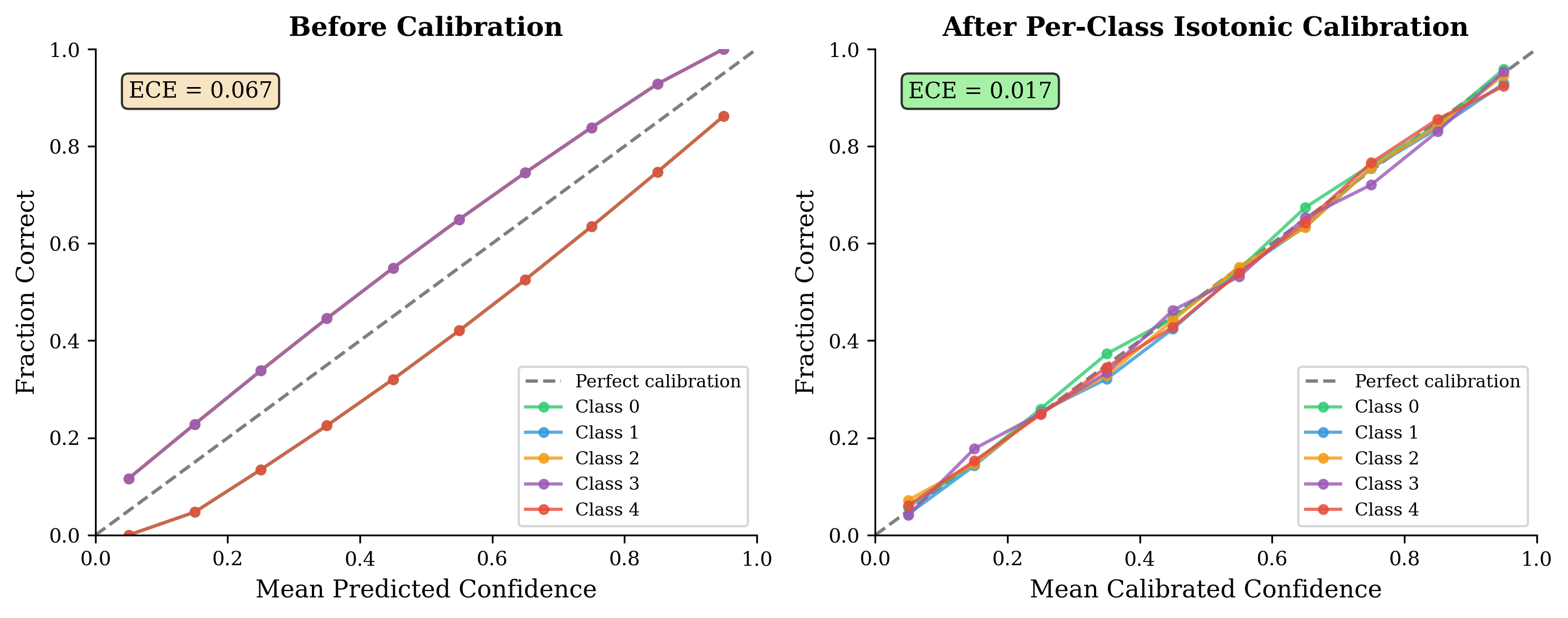}
\caption{Calibration curves before and after per-class isotonic calibration.}
\label{fig:calibration}
\end{figure}

We learn separate isotonic calibration functions per predicted class, following Zadrozny and Elkan \cite{zadrozny2002transforming}:
\begin{equation}
p^*_c = f_c(\hat{p}) \quad \text{where } f_c: [0,1] \to [0,1]
\end{equation}
is monotonically non-decreasing.
For each model $m$ and class $c$:
(a) we collect predictions where $\hat{y}_i = c$,
(b) compute correctness $z_i = \mathbf{1}[\hat{y}_i = y_i]$, and
(c) fit $f_c = \text{IsotonicRegression}(\hat{p}, z)$.
We choose isotonic regression over Dirichlet calibration \cite{kull2019beyond} for
(a) its nonparametric flexibility, and (b) monotonicity preservation for threshold-based routing.
Figure~\ref{fig:calibration} shows calibration curves before and after per-class calibration.

\subsection{Cascade Decision Logic}

\begin{algorithm}[t]
\small
\caption{Calibrated Cascade Inference}
\label{alg:cascaded_inference}
\begin{algorithmic}[1]
\REQUIRE Query $q$, title $t$, thresholds $\{\tau_m\}$
\FOR{model $m \in$ [CrossEncoder, Gemma, LLaMA]}
    \STATE $(\hat{y}_m, \hat{p}_m) \leftarrow m(q, t)$
    \STATE $p^*_m \leftarrow f_{m,\hat{y}_m}(\hat{p}_m)$ \hfill \textit{\{Per-class calibration\}}
    \IF{$p^*_m \geq \tau_m$}
        \RETURN $\hat{y}_m$
    \ENDIF
\ENDFOR
\STATE $\hat{y} \leftarrow \text{MajorityVote}(\hat{y}_1, \hat{y}_2, \hat{y}_3)$
\STATE \textbf{if} tie \textbf{then} $\hat{y} \leftarrow \max(\hat{y}_1, \hat{y}_2, \hat{y}_3)$
\RETURN $\hat{y}$
\end{algorithmic}
\end{algorithm}

Given calibrated confidence thresholds $\tau_1, \tau_2, \tau_3$ for each model, the cascade proceeds as mentioned in the Algorithm \ref{alg:cascaded_inference}.

The ensemble fallback uses majority voting with a max-prediction tiebreaker that biases toward higher relevance---appropriate for offline use cases where over-penalizing relevant ads is costlier than mild over-scoring.

\subsection{Implementation Details}

How are the cascade models trained?
Table~\ref{tab:hparams} reports training hyperparameters for the three classifiers.
We trained on NVIDIA H100 GPUs and benchmarked inference latency on A100 (40GB) for production parity.
The Cross-Encoder fine-tunes all 184M parameters; Gemma-2B and LLaMA-8B use LoRA \cite{hu2022lora} on attention projections together with a 5-class classification head.
All models use AdamW with linear warmup (10\% of total steps) followed by linear decay, and early stopping on validation cross-entropy.

\begin{table}[t]
\centering
\small
\begin{tabular}{lccc}
\toprule
\textbf{Hyperparameter} & \textbf{Cross-Enc.} & \textbf{Gemma-2B} & \textbf{LLaMA-8B} \\
\midrule
Learning rate     & 1e-5  & 5e-4  & 5e-4  \\
Optimizer         & AdamW & AdamW & AdamW \\
Max epochs        & 20    & 5     & 5     \\
Batch size        & 128   & 4     & 4     \\
Warmup ratio      & 0.1   & 0.1   & 0.1   \\
LR schedule       & linear & linear & linear \\
LoRA $r$, $\alpha$ & ---  & 16, 16 & 16, 16 \\
Train (H100-hrs)  & ${\sim}8$  & ${\sim}16$ & ${\sim}68$ \\
\bottomrule
\end{tabular}
\caption{Training hyperparameters and single-H100 wall-clock training cost for the three cascade models.}
\label{tab:hparams}
\end{table}

\section{Experiments and Evaluation}

\subsection{Research Questions}

We structure our evaluation around four questions:
\textbf{Q1:} How do fine-tuned classifiers compare to off-the-shelf LLMs?
\textbf{Q2:} Does per-class calibration improve cascade routing over global methods?
\textbf{Q3:} What is the cost-accuracy tradeoff of the cascade versus single-model baselines?
\textbf{Q4:} How do fine-tuning, cascading, and calibration each contribute to the final accuracy and cost?

\subsection{Dataset and Setup}

We fine-tune on 1.2M query-product pairs from a major e-commerce platform's sponsored search, collected over 18 months.
We evaluate on a held-out set of 100K pairs annotated by trained raters on a 5-point scale (0=Embarrassing through 4=Excellent and class distribution approximately balanced at 18--22\% per class).
The evaluation set is split: 50\% for calibration fitting, and 50\% for final evaluation.
All metrics use bootstrap CIs (1,000 resamples) and differences $>$0.5\% are significant at $p < 0.01$.

\subsection{Individual Model Performance (Q1)}

We showed our model performance comparison with 95\% bootstrap confidence intervals in Table~\ref{tab:individual} .

\begin{table}[t]
\centering
\small
\begin{tabular}{lcc}
\toprule
\textbf{Model} & \textbf{Accuracy (95\% CI)} & \textbf{Latency} \\
\midrule
GPT-4 (off-the-shelf) & 68.2 $\pm$ 0.9\% & 500ms \\
Claude-3 (off-the-shelf) & 70.1 $\pm$ 0.9\% & 450ms \\
\midrule
Cross-Encoder (ours) & 87.5 $\pm$ 0.6\% & 5ms \\
Gemma-2B (ours) & 88.2 $\pm$ 0.6\% & 50ms \\
LLaMA-8B (ours) & 88.9 $\pm$ 0.6\% & 200ms \\
\bottomrule
\end{tabular}
\caption{Accuracy and latency of individual models.}
\label{tab:individual}
\end{table}

\noindent\textbf{Observation 1: Fine-tuning closes a 20-point accuracy gap.}
All fine-tuned models achieve 87--89\% vs.\ 68--70\% for off-the-shelf LLMs.
Even the smallest Cross-Encoder (184M params) outperforms GPT-4 by 20 points while being 100$\times$ faster.
We argue that classification heads produce well-defined probability distributions, whereas generative prompting is inherently fragile and poorly calibrated.

\begin{figure}[t]
\centering
\includegraphics[width=0.6\columnwidth]{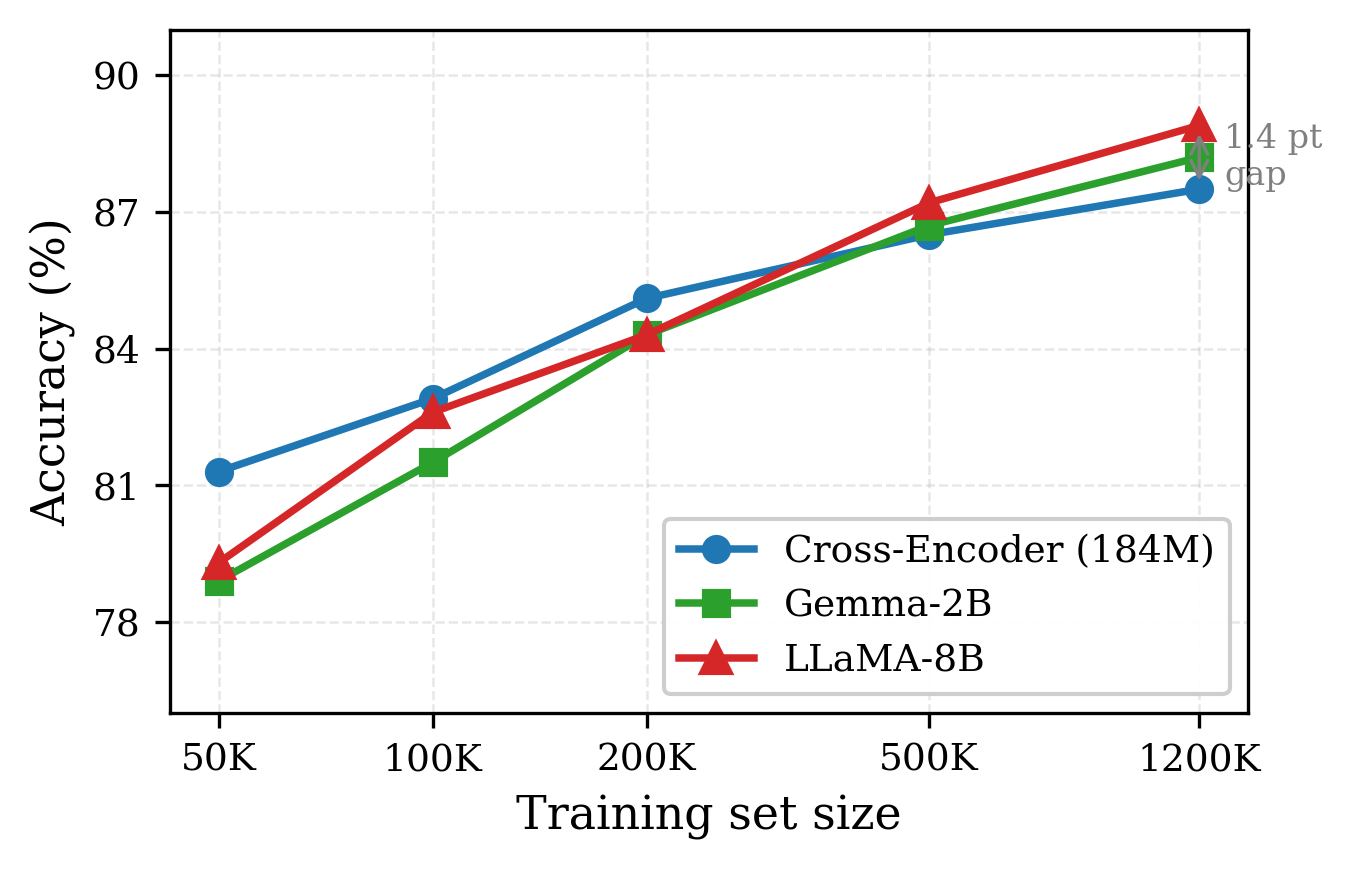}
\caption{Learning curves by training-set size; smaller models saturate earlier while larger models keep improving.}
\label{fig:learning}
\end{figure}

\noindent\textbf{Observation 1b: Smaller models saturate earlier.}
Figure~\ref{fig:learning} reports accuracy as a function of training-set size for each model.
The Cross-Encoder gains only 6.2 points from 50K to 1.2M examples (81.3$\to$87.5), while LLaMA-8B gains 9.6 points (79.3$\to$88.9) and Gemma-2B gains 9.3 points (78.9$\to$88.2).
The pattern is consistent with capacity-driven scaling: smaller models reach their plateau with modest data, while larger models continue to benefit.
This sample-efficiency gradient justifies the cascade ordering---the Cross-Encoder is a reliable cheap first stage, while the larger models earn their inference cost on the harder long tail.

\subsection{Calibration Method Comparison (Q2)}

We presented our calibration methods comparison in Table~\ref{tab:calibration}.

\begin{table}[t]
\centering
\small
\begin{tabular}{lccc}
\toprule
\textbf{Method} & \textbf{ECE}$\downarrow$ & \textbf{Casc.\ Acc.} & \textbf{Bal.\ Acc.} \\
\midrule
No calibration & .067$\pm$.008 & 88.2$\pm$0.6 & 88.0$\pm$0.6 \\
Temperature & .040$\pm$.005 & 88.5$\pm$0.6 & 88.3$\pm$0.6 \\
Platt scaling & .027$\pm$.004 & 88.5$\pm$0.6 & 88.3$\pm$0.6 \\
Histogram bin. & .010$\pm$.003 & 88.5$\pm$0.6 & 88.3$\pm$0.6 \\
Isotonic (global) & .012$\pm$.003 & 88.5$\pm$0.6 & 88.3$\pm$0.6 \\
\textbf{Iso.\ (per-class)} & .017$\pm$.004 & \textbf{89.1$\pm$0.6} & \textbf{89.0$\pm$0.6} \\
\bottomrule
\end{tabular}
\caption{Calibration comparison for cascade routing; per-class isotonic gives the best cascade accuracy.}
\label{tab:calibration}
\end{table}

\noindent\textbf{Observation 2: Per-class calibration yields the best cascade accuracy despite higher global ECE.}
The most stringent comparison is against the strongest calibration baseline---global isotonic regression---rather than against the uncalibrated cascade.
Per-class isotonic improves cascade accuracy from 88.5\% to 89.1\% over global isotonic (+0.6 points, $p < 0.05$, paired bootstrap), and from 88.2\% to 89.1\% over no calibration (+0.9 points).
Global ECE (0.017 for per-class vs.\ 0.010 for histogram binning) masks the class-conditional improvements that matter for routing (see Figure~\ref{fig:calibration}).

\subsection{Cascade Performance (Q3)}

We illustrated cascaded performance across threshold configurations in Table~\ref{tab:cascade}.

\begin{table}[t]
\centering
\small
\begin{tabular}{cccrr}
\toprule
\textbf{XE} & \textbf{Gem} & \textbf{Lla} & \textbf{Acc.\ (95\% CI)} & \textbf{Savings} \\
\midrule
0.92 & 0.98 & 0.90 & 88.9 $\pm$ 0.6\% & 46.8\% \\
0.90 & 0.90 & 0.90 & 89.0 $\pm$ 0.6\% & 44.2\% \\
0.85 & 0.85 & 0.85 & 89.1 $\pm$ 0.6\% & 39.8\% \\
\textbf{0.85} & \textbf{0.80} & \textbf{0.75} & \textbf{89.1 $\pm$ 0.6\%} & \textbf{50.1\%} \\
\bottomrule
\end{tabular}
\caption{Representative thresholded cascade configurations and their cost savings.}
\label{tab:cascade}
\end{table}

\noindent\textbf{Observation 3: The cascade achieves ensemble-level accuracy at half the cost.}
The production configuration achieves 89.1\% accuracy with 50.1\% cost savings.
For reference, the full ensemble (all three models) achieves 89.3\% at 1.28$\times$ cost, while LLaMA-8B alone achieves 88.9\% at 1.0$\times$. The cascade matches ensemble accuracy at lower cost and lower latency.

\noindent\textbf{Observation 4: The Cross-Encoder resolves 74.5\% of queries at higher-than-standalone accuracy.}
Table~\ref{tab:sources} shows prediction origins.
The Cross-Encoder handles 74.5\% of queries at 91.2\% which is well above its standalone 87.5\% ($p < 0.001$), because calibration routes only high-confidence predictions to the cheap first stage.
The ensemble fallback handles 7.1\% at 79.2\%, acceptable for inherently ambiguous samples.

\begin{table}[t]
\centering
\small
\begin{tabular}{lrc}
\toprule
\textbf{Decision Source} & \textbf{Traffic} & \textbf{Accuracy (95\% CI)} \\
\midrule
Cross-Encoder (confident) & 74.5\% & 91.2 $\pm$ 0.7\% \\
Gemma (confident) & 12.5\% & 86.8 $\pm$ 1.9\% \\
LLaMA (confident) & 5.9\% & 84.5 $\pm$ 2.9\% \\
Ensemble fallback & 7.1\% & 79.2 $\pm$ 3.0\% \\
\midrule
\textbf{Overall} & 100\% & \textbf{89.1 $\pm$ 0.6\%} \\
\bottomrule
\end{tabular}
\caption{Traffic and accuracy by decision source in the production cascade.}
\label{tab:sources}
\end{table}

\begin{table}[!htb]
\centering
\small
\begin{tabular}{lrr}
\toprule
\textbf{Configuration} & \textbf{Acc.} & \textbf{Rel.\ Cost} \\
\midrule
\multicolumn{3}{l}{\textit{Off-the-shelf LLMs (API)}} \\
GPT-4 (single)               & 68.2\% & ---  \\
GPT-4 + Claude-3 (cascade)   & 71.4\% & ---  \\
\midrule
\multicolumn{3}{l}{\textit{Fine-tuned (ours, local A100)}} \\
LLaMA-8B (single)            & 88.9\% & 1.00$\times$ \\
Cascade, uncalibrated        & 88.2\% & 0.50$\times$ \\
Cascade, global isotonic     & 88.5\% & 0.50$\times$ \\
\textbf{Cascade, per-class iso.} & \textbf{89.1\%} & \textbf{0.50$\times$} \\
\bottomrule
\end{tabular}
\caption{Ablation of fine-tuning, cascading, and calibration. Relative cost is normalized to fine-tuned LLaMA-8B.}
\label{tab:ablation}
\end{table}

\subsection{Decomposing the Gains (Q4)}

How much of our 89.1\% accuracy comes from fine-tuning versus cascading versus calibration?
Table~\ref{tab:ablation} isolates each component's contribution.

\noindent\textbf{Observation 5: Fine-tuning contributes the bulk of the accuracy gain.}
A single fine-tuned LLaMA-8B classifier (88.9\%) closes 20.7 points of the gap over GPT-4 (68.2\%).
By contrast, cascading off-the-shelf LLMs (GPT-4 $\rightarrow$ Claude-3) gains only 3.2 points over GPT-4 alone (68.2$\to$71.4\%), consistent with the cascade-bound argument of Jitkrittum et al.\ \cite{jitkrittum2023when}: routing cannot exceed the accuracy of its constituents by much.

\noindent\textbf{Observation 6: Cascading delivers cost reduction without sacrificing accuracy.}
Moving from a single fine-tuned LLaMA-8B (88.9\%, 1.00$\times$) to an uncalibrated cascade gives 88.2\% at 0.50$\times$ cost---a 0.7-point drop for 50\% cost savings.
Calibration recovers and then exceeds the single-model accuracy.

\noindent\textbf{Observation 7: Calibration contributes a small but consistent gain.}
Per-class isotonic calibration adds 0.9 points over the uncalibrated cascade and 0.6 points over the strongest calibration baseline (global isotonic), both statistically significant ($p < 0.05$).

The decomposition supports our central claim: \textit{accuracy and cost are orthogonal optimizations}---fine-tuning drives accuracy, cascading drives cost, and calibration provides a small additive gain that is reliable across deployment windows.

\section{Production Deployment and Discussion}

How does the system perform in real-world offline workflows?
Since deployment in Q3 2024, the cascade has processed over 150 million annotations across six offline use cases:
(a) training data preparation (100M+ pairs),
(b) NDCG evaluation (10K+ annotations per experiment),
(c) root cause analysis (1-2M annotations within hours),
(d) guardrail calibration,
(e) feature evaluation via A/B tests, and
(f) ANN dictionary preparation.
All use cases are \textit{fully offline}. The cascade is never in the serving path.
Median turnaround dropped from 5 days to 1-3 hours.  Three teams actively use the system across Advertising, Marketing, and Recommendations.

We discuss several practical insights:

\noindent\textbf{a. Does input format affect accuracy?}
Yes, dramatically.
Using the exact template from training is critical and even minor variations degrade accuracy by 10--15\%.

\noindent\textbf{b. How fresh must calibrators be?}
We retrain quarterly using a rolling 50K-sample window.
Stale calibrators lead to suboptimal routing that silently degrades accuracy.

\noindent\textbf{c. Are high thresholds always better?}
High thresholds (e.g., 0.98) render intermediate models useless.
Lower thresholds (0.80--0.85) better utilize the full cascade.

\noindent\textbf{d. How does our system compare to related work?}
A comparable study fine-tuned LLaMA-2 7B for 3-class sponsored search relevance, achieving 89.43\% \cite{walmart2024relevance}.
Our system achieves nearly identical accuracy on a harder 5-class task while halving compute which suggests cascade architecture provides efficiency gains beyond single-model approaches.

\noindent\textbf{Limitations.}
Our evaluation covers a single domain (sponsored search), so the 20-point accuracy gain reflects fine-tuning in this setting rather than cascade architecture alone.
The system also requires substantial labeled data (1.2M examples for fine-tuning and 50K for calibration).
Latency comparisons use local A100 inference vs.\ cloud APIs, although our primary efficiency claim (50\% savings vs.\ running LLaMA-8B on every query) is deployment-independent.
Finally, proprietary data prevents external reproduction, though we provide detailed methodology for replication on other datasets.

\section{Conclusion}

We presented a calibrated model cascade for cost-efficient offline relevance annotation.
The core insight is that \textbf{accuracy and cost are orthogonal optimizations}: fine-tuning contributes ${\sim}20$ accuracy points, cascading halves compute cost without sacrificing accuracy, and per-class isotonic calibration adds a small but reliable gain (+0.6 points over the strongest baseline) on top.
The system has been deployed in production, processing 150M+ annotations across six offline use cases.
We argue that our work is a building block for scalable offline annotation pipelines, with natural extensions including intra-model early exits \cite{schuster2022confident}, human-in-the-loop deferral, and adaptive model orderings.

% \section*{Ethics Statement}

% This system is designed exclusively for offline annotation. It does not replace human judgment for user-facing decisions.
% All training annotations were collected under standard employment conditions.
% The max-prediction tiebreaker biases toward higher relevance; we monitor via weekly audits against human spot-checks.

\section*{Declaration on Generative AI}
During the preparation of this work, the author(s) used Generative AI only for formatting assistance in plotting code used to produce figures, and not for drafting, analysis, interpretation, or any other part of the manuscript. The author(s) reviewed, corrected, and validated the generated code and figures and take(s) full responsibility for the publication's content.

% If no Generative AI tools were employed:
% The author(s) have not employed any Generative AI tools.
%
% If Generative AI tools were employed, use the CEUR-WS activity taxonomy:
% During the preparation of this work, the author(s) used ...

\bibliography{references}

\end{document}